\documentclass[twoside]{article}
\usepackage{fleqn,espcrc2}

\usepackage{graphicx}
\usepackage[figuresright]{rotating}

\newcommand{\AmS}{{\protect\the\textfont2
  A\kern-.1667em\lower.5ex\hbox{M}\kern-.125emS}}

\title{Supernova Neutrino Detection}

\author{Kate Scholberg\address{Boston University Dept. of Physics,
        590 Commonwealth Ave., Boston, MA 02215}}%

\begin{document}

\begin{abstract}
  World-wide, several detectors currently running or nearing
  completion are sensitive to a core collapse supernova neutrino
  signal in the Galaxy.  I will briefly describe the nature of the
  neutrino signal and then survey current and future detection 
  techniques.  I will
  also explore what physics and astrophysics
  we can learn from the next Galactic core collapse.

\vspace{1pc}
\end{abstract}

\maketitle

\section{THE NEUTRINO SIGNAL}

Astronomers expect that every 30 years or so, a massive star in our
Galaxy will reach the end of its life.  When the core of such a star
collapses, it emits nearly all of the gravitational binding energy of
a neutron star in the form of neutrinos, some
$E_b~\sim~3\times10^{53}$~ergs.  Less than 1\% of this energy is
released in the form of kinetic energy and optically visible
radiation. The remainder is radiated in neutrinos, of which
approximately 1\% is $\nu_e$ from an initial ``breakout'' burst
and the remaining 99\% are $\nu \bar{\nu}$ pairs from the later
cooling reactions, roughly equally distributed among flavors.  Average
neutrino energies will be about 12~MeV for $\nu_e$, 15~MeV for
$\bar{\nu}_e$, and 18~MeV for all other flavors.  This hierarchy of
energies is explained by the fact $\bar{\nu}_e$'s have fewer charged
current (CC) interactions with the neutron-rich
stellar core than do $\nu_e$'s;
they decouple deeper inside the core where it is hotter
and so emerge with higher average energies than do $\nu_e$'s.
Similarly, $\nu_{\mu,\tau}$\footnote{I will use the
symbol $\nu_{\mu,\tau}$ to refer to muon and/or tau neutrinos
and their antineutrinos.}, which have only neutral current (NC)
interactions with the core's matter, emerge with even higher energies.
The neutrinos are emitted over a total timescale of tens of
seconds, with about half emitted during the first 1-2~seconds, and
with the spectrum eventually softening as the proto-neutron star
cools.  
Burrows~\cite{Burrows} sketches the expected neutrino signal. Some
more recent developments in core collapse theory are described in
these proceedings by Mezzacappa~\cite{Mezzacappa} and
Prakash~\cite{Prakash}.  

\section{DETECTORS}

Several kinds of detector are capable of detecting a burst of
neutrinos from a gravitational collapse in our Galaxy.
Most of the detector types described here actually have primary
purposes other than supernova neutrino detection, 
{\it e.g.} proton decay searches, solar and
atmospheric neutrino physics, neutrino oscillation
studies, and astrophysical neutrino source searches.

\subsection{Scintillation Detectors}

The hydrogen in hydrocarbon scintillator has a high
cross-section for the charged current (CC) antineutrino absorption
reaction,
\begin{equation}\label{eq:invbetadecay}
\bar{\nu}_e~+~p~\rightarrow~e^+~+~n.
\end{equation}

The positron energy is highly correlated with the neutrino energy.
The neutron from reaction~\ref{eq:invbetadecay} may also be detectable
via the time-delayed fusion of the neutron with protons in the
scintillator:
\begin{equation}
n~+~p~\rightarrow~d~+~\gamma~(2.2~\rm{MeV}),
\end{equation}
where the $\gamma$ can be detected by its Compton
scattering.  The average moderation time for the neutron is
$\sim$10~$\mu$s, and the average capture time is $170~\mu$s; the
neutron typically wanders for tens of centimeters before being
captured.  Therefore, a time-delayed coincidence of a $\sim$10~MeV
pulse with a $\sim$2~MeV signal can provide a signature for the interaction
of supernova $\bar{\nu}_e$ in liquid scintillator detectors.

Scintillators are also sensitive to 
CC and NC neutrino-electron scattering and the superallowed NC excitation of
$^{12}$C by neutrinos of all flavors, 
although cross-sections are much lower than
for inverse beta decay.  The NC $^{12}$C excitation can
be detected (and, to some extent, be tagged) via 
the 15~MeV de-excitation gamma; such events should represent about 10\%
of the total number of interactions, although detection efficiency
varies, depending on detector geometry.

Scintillator detectors are not generally able to point
to a supernova source.  The positron products of the
predominant inverse beta decay reaction are emitted nearly
isotropically, and the scintillation light is isotropized.
However the neutron emission is asymmetric and if
the relative positions of the positron and the neutron
absorption can be measured, some directional information can be
extracted.  The CHOOZ experiment~\cite{chooz} has recently 
demonstrated of this principle using reactor neutrinos.

Currently, the MACRO~\cite{GCmacro} and LVD~\cite{LVDref} detectors at
Gran Sasso have enough scintillator mass to be sensitive to a
gravitational collapse anywhere in the Galaxy.  LSD and Baksan, with
less mass and sensitivity, are also still running.  Scintillation
detectors which will come online over the next several years include
Borexino at the Gran Sasso and KamLAND in Japan.

\subsection{Water Cherenkov Detectors}


The charged current inverse beta decay
reaction, reaction~\ref{eq:invbetadecay}, will dominate the supernova
neutrino interaction rate in water Cherenkov detectors.   Therefore, as
for scintillator detectors, the primary sensitivity is to the
$\bar{\nu}_e$ component of supernova neutrino radiation.  Light water
Cherenkov detectors are also capable of 
detecting CC and NC
neutrino-electron scattering reactions:
\begin{equation}\label{eq:scatt}
\nu_x~+~e^-~\rightarrow~\nu_x~+~e^-,
\end{equation}
where $\nu_x$ represents a neutrino of any flavor.
Although elastic scattering events should represent only a few percent
of the total supernova signal,
this reaction has potential for allowing
reconstruction of the direction to the supernova source, because
the $e^-$ direction, which can
be determined from the Cherenkov cone direction,
is related to the $\nu_e$ direction.

Some NC interactions of neutrinos
on oxygen are expected,
\begin{equation}
^{16}{\rm O}~+~\nu_x~\rightarrow~^{16}{\rm O}^{*}~+~\nu_x,
\end{equation}
which can be detected via a cascade of 5-10~MeV de-excitation gammas;
these may be useful for neutrino mass measurements (see
section~\ref{nuphys}).  Finally, a few percent of interactions will be
CC interactions of $\nu_e$ and $\bar{\nu}_e$ on $^{16}$O and $^{18}$O.

The largest existing water Cherenkov detector is the Super-Kamiokande
detector~\cite{superk} in Japan, which should observe thousands of neutrinos
from a Galactic supernova.

\subsection{Heavy Water Detectors}

Heavy water neutrino detectors exploit
the relatively high cross-sections for the interaction of neutrinos
with deuterons.  Both CC and NC
deuteron breakup reactions are detectable, via Cherenkov light 
of the lepton products for the CC case, and via neutron detection
of some sort for the NC and $\bar{\nu}_e$ CC cases.
Two classes of reactions are relevant here:

\begin{enumerate}
\item The neutral current interaction
of neutrinos with deuterons:
\begin{equation}
\nu_x~+~d~\rightarrow~n~+~p~+~\nu_x.
\end{equation}
In this case the neutron is detected in some way, 
via Cherenkov light from Compton-scattered
gamma rays produced from absorption of the neutrons on nuclei
in the detector medium, or 
using dedicated
neutron detectors.
The proton is below Cherenkov threshold and is therefore invisible.
No energy or neutrino direction
is available for this reaction; however, the reaction is sensitive
to all flavors of neutrinos.

\item The charged current interaction of electron neutrinos and
anti-neutrinos with deuterons:
\begin{equation}\label{eq:ccbreakup1}
\nu_e~+~d~\rightarrow~p~+~p~+~e^-
\end{equation}
\begin{equation}\label{eq:ccbreakup2}
\bar{\nu}_e~+~d~\rightarrow~n~+~n~+~e^+.
\end{equation}

In these cases, the $e^+$ or $e^-$ can be detected by its
Cherenkov light, and energy and directional information can be
extracted.  In addition, the neutrons from the anti-neutrino reaction
can be detected with a time-delayed coincidence (also tagging the
anti-neutrino reaction).  The products of these reactions retain some
information about the direction of the incoming neutrino.

\end{enumerate}

The neutral current (NC) sensitivity of heavy water will make its supernova
signal especially rich.  In
addition to providing a unique diagnostic of supernova processes, a NC
detector may be able to provide strong constraints (or measurement) of
absolute neutrino mass from relative time of flight measurements (see
section~\ref{nuphys}) in conjunction with CC data from 
the same detector, or with data from other detectors with
primarily $\bar{\nu}_e$ flavor sensitivity.

At present, the only example of a heavy water supernova neutrino
detector is the SNO~\cite{SNO} detector in Canada.

\subsection{Long String Water Cherenkov Detectors}

Long string water Cherenkov detectors consist of long strings of
photomultiplier tubes (PMTs) hanging in very clear water or buried in ice.  These
detector arrays are designed primarily to observe high energy ($>>$GeV)
neutrinos by looking for Cherenkov light from upward-going
neutrino-induced muons.
However, these detectors have some supernova neutrino sensitivity as well.
Although 
the PMTs in the array are too sparse to
allow event-by-event supernova neutrino reconstruction at low energy,
a large burst of supernova neutrinos will produce enough
Cherenkov photons from interactions in the ice surrounding the
PMTs to cause a coincident increase in single PMT count rates from
many PMTs~\cite{Amanda}. For instance, the track 
length for 20~MeV positrons in ice is expected to be about 10~cm, 
from which roughly
3000 Cherenkov photons are emitted~\cite{Amanda2}; an absorption
length of $\sim$100~m then leads to an effective volume of $\sim$400~m$^3$ 
per PMT. The supernova sensitivity of such a long string
detector is strongly dependent on the noisiness of its phototubes and
ambient light background.  Again, since reaction~\ref{eq:invbetadecay}
dominates, primary sensitivity is to $\bar{\nu}_e$. There is no
direction sensitivity, although in principle participation in a
triangulation measurement using timing is
possible.

Examples of long string Cherenkov detectors are AMANDA in the
Antarctic ice, the Lake Baikal detector, and NESTOR and ANTARES in the
Mediterranean.  Currently the most promising for supernova detection
in terms of a low background noise rate is the AMANDA detector~\cite{Amanda2}.

\subsection{High Z/Neutron Detectors}\label{highz}

High Z/neutron detectors are primarily sensitive to the high energy
component of the supernova neutrino flux.  The idea is to
observe neutrons emitted from the NC reaction
\begin{equation}
\nu_x~+~(A,Z)~\rightarrow~(A-1,Z)~+~n~+~\nu_x.
\end{equation}
In addition, neutrons can arise
from CC neutrino interactions\begin{equation}
\nu_e~+~(A,Z)~\rightarrow~(A-1,Z+1)~+~n~+~e^-,
\end{equation}
\begin{equation}
\bar{\nu}_e~+~(A,Z)~\rightarrow~(A-1,Z-1)~+~n~+~e^+;
\end{equation}

The $e^+$ or $e^-$ may be detectable as well.
These reactions occur primarily for high energy neutrinos; therefore the
neutrino flavors detected are expected to be mostly muon and tau (which are
emitted with higher temperatures).  Just as for the other detectors with
NC sensitivity, multi-flavor sensitivity 
is useful for probing core collapse physics,
and will be especially valuable for neutrino mass studies via time delay
measurements~\cite{Cline}.  High Z materials such as lead or iron are appropriate, since
the cross-section increases along with number of nucleons in the nucleus, and
a large quantity of relatively pure material can be obtained cheaply.
A combination of different materials can give rough neutrino
spectral information due to different energy dependence of cross-sections for
different nuclei; this in turn can yield information about neutrino
oscillation, which would produce distortion of the energy spectra.
A particularly promising target material for
  oscillation tests is $^{208}$Pb~\cite{Fuller}.

Proposed detectors of this type include OMNIS and LAND.

\subsection{Other Detectors}

Some other supernova-sensitive detectors (existing, 
under construction, or proposed) are:

\begin{itemize}
\item ICANOE (Imaging and CAlorimetric Neutrino Oscillation Experiment): 
liquid argon drift chambers are
to be built at the Gran
Sasso Laboratory. These should have sensitivity to $\nu_e$
from supernovae via
\begin{equation}
\nu_e~+^{40}\rm{Ar}~\rightarrow~^{40}\rm{K}^{*}~+~e^{-}.
\end{equation}

The final state electron is detected, and in addition
the de-excitation gamma rays from $^{40}\rm{K}^{*}$
can be detected.

\item Radiochemical detectors: the $^{37}$Cl solar neutrino detector
at the Homestake mine, the gallium solar neutrino detectors such as
GALLEX and SAGE, or a possible $^{127}$I detector, will register
counts if supernova neutrinos interact inside them.  Since the
sensitive chemicals are only extracted at very long time intervals
($\sim$ weeks or months), these detectors have no time (or energy)
resolution and are of limited value for a stellar collapse search.
However, if a burst of gravitational collapse neutrinos were confirmed
in one or many of the real-time neutrino detectors, radiochemical
experiments could perform prompt extractions to determine whether any
counts over background were registered during the relevant time
period.  Furthermore, recent proposals for quasi-real-time triggered
chlorine extractions
(\cite{Lande}) may make radiochemical technology more interesting
for supernova neutrinos.

\item Gravitational wave detectors: large interferometer experiments
  under construction such as LIGO and VIRGO will have the capability
  of detecting gravitational wave signals from asymmetric supernova
  explosions (although the details of a stellar collapse gravitational
  wave signal are not yet well understood).  The gravitational wave
  signal may be more prompt even than the neutrino signal.


\end{itemize}

\subsection{Detector Summary}

Table~\ref{tab:detector_types} gives a brief overview of the
neutrino detector types mentioned here.

\begin{table*}[htb]
\caption{Supernova neutrino detector types and their capabilities.\label{tab:detector_types}}
\newcommand{\m}{\hphantom{$-$}}
\newcommand{\cc}[1]{\multicolumn{1}{c}{#1}}
\renewcommand{\tabcolsep}{1.1pc} 
\renewcommand{\arraystretch}{1.0} 
\begin{tabular}{@{}llllll}
\hline
Detector type & Material & Energy & Time & Point & Flavor \\ 
\hline
scintillator & C,H & y & y & n & $\bar{\nu}_e$\\ 
water Cherenkov &  H$_2$0 & y & y & y & $\bar{\nu}_e$ \\ 
heavy water & D$_2$0 & NC: n & y & n & all \\ 
         &      & CC: y & y & y & $\nu_e$,$\bar{\nu}_e$\\ 
long string water Cherenkov& H$_2$O & n & y & n & $\bar{\nu}_e$ \\ 
liquid argon & Ar & y & y & y & $\nu_e$ \\ 
high Z/neutron  & NaCl, Pb, Fe & n & y & n & all \\ 
radiochemical & $^{37}$Cl, $^{127}$I, $^{71}$Ga  & n & n & n & $\nu_e$ \\ 
\hline
\end{tabular}\\[2pt]
\end{table*}

In summary, scintillator and water Cherenkov detectors are sensitive
primarily to $\bar{\nu}_e$; detectors with NC
capabilities (heavy water, high Z/neutron) are sensitive to all
flavors.  Water Cherenkov and heavy water detectors have significant
pointing capabilities.  All can see neutrinos in real-time, except
radiochemical. All have good energy resolution, except long string
water Cherenkov, high Z and radiochemical.

Table~\ref{tab:specific_detectors} lists specific supernova
neutrino detectors and their capabilities.  

\begin{table*}[htb]
\caption{Specific supernova neutrino detectors.}\label{tab:specific_detectors}
\newcommand{\m}{\hphantom{$-$}}
\newcommand{\cc}[1]{\multicolumn{1}{c}{#1}}
\renewcommand{\tabcolsep}{1.0pc} 
\renewcommand{\arraystretch}{1.0} 
\begin{tabular}{@{}llllll}
\hline
Detector & Type & Mass & Location  & \# of events & Status\\ 
         &      & (kton) &         & @8.5 kpc &  \\ 
\hline
Super-K & H$_2$O Ch. & 32    & Japan  & 5000 & running \\ 
MACRO & scint.& 0.6 & Italy & 150 &  running \\ 
SNO & H$_2$O,& 1.4 & Canada &  300 & running \\
    & D$_2$O & 1   &        &  450 &  \\ 
LVD & scint. & 0.7 & Italy & 170 & running\\ 
AMANDA & long string  & M${\rm eff}\sim$0.4/pmt & Antarctica & & running \\
Baksan & scint. & 0.33  & Russia  & 50 &  running\\ 
Borexino & scint. & 0.3  & Italy  & 100 &  2001\\ 
KamLAND & scint. & 1  &  Japan  & 300 &  2001\\ 
OMNIS  & high Z (Pb/Fe) & 10(Fe)+4(Pb) &  USA & 2000 &  proposed \\ 
LAND  & high Z (Pb)  & &  Canada &  &  proposed \\ 
Icanoe & liquid argon & 9 & Italy & & 2005 \\ \hline
\end{tabular}
\end{table*}

\section{WHAT CAN WE LEARN?}\label{nuphys}

\subsection{Overview}

A comprehensive review of all that might be learned  
from a Galactic signal would be a daunting task, so 
I will give only a brief overview, and then will focus on the one
subject of absolute neutrino mass sensitivity via time of flight.

\subsubsection{Neutrino Physics} A core collapse event provides a powerful and
  distant natural source of neutrinos, allowing the study of 
  neutrino properties, such as mass,
  flavor oscillations, and charge.  Some of the most interesting
  possible measurements are of absolute neutrino mass, exploiting the
  time of flight delays over the long distance from the supernova: one
  can look for an energy-dependent time spread in the observed events,
  or a flavor-dependent delay.  In addition, the effects of neutrino
  oscillation in the core of the star manifest themselves as an
  anomalously hot $\nu_e$ (or $\bar{\nu}_e$) spectrum (e.g.~\cite{Fuller,Dighe}).  More information can be extracted: 
the neutrino data from SN1987A set limits on neutrino
  lifetime, charge, number of neutrino families, neutrino magnetic
  moment~\cite{Schramm}, and even extra dimensions~\cite{Arkani-Hamed}.
Although there will always be at least some collapse
model dependence in any conclusions,
in many cases it is possible to make quite robust assumptions about the neutrino
source.

\subsubsection{Core Collapse Physics} The time structure, energy spectrum
  and flavor composition of the neutrino burst will yield information
  on astrophysical stellar collapse processes, even for ``silent''
  collapses, i.e. those occurring without strong electromagnetic
  fireworks or in regions of the Galaxy obscured by dust.  The
  neutrino signal will help to understand the explosion mechanism and
  the mechanisms of proto neutron star cooling (see
  ~\cite{Mezzacappa,Prakash}).  In addition, the formation of a
  black hole may be signaled by a sharp cutoff in neutrino
  luminosity~\cite{Burrows}.

\subsubsection{Astronomy from an Early Alert}
The neutrino burst produced by the core collapse emerges promptly from
the stellar envelope.  However, the shock wave produced by the
collapse takes some time to travel outwards.
The time of first shock breakout
is highly dependent on the nature of the stellar envelope,
and can range from minutes for bare-core stars to hours for red
giants.  For SN1987A, first light was observed about 2.5 hours after
the neutrino burst; the first observable photons probably reached us
about one hour earlier than that.
The observation of very early light from a supernova just after shock
breakout is astrophysically very interesting~\cite{HST}, and
rare for extragalactic supernovae.  One can learn about the 
supernova progenitor
and its immediate environment.
And of course, an observation of very early supernova light could also
yield entirely unexpected effects.  
It is possible that a core collapse event will not yield an
optically bright supernova, either because the explosion ``fizzles'',
or because the supernova is in an optically obscured region of
the sky.  In the latter case there may still be an observable
event in some wavelengths, or in gravitational radiation.

\subsection{Absolute Neutrino Mass}

The time delay for a neutrino of energy $E$ and mass $m_\nu$ is given by
$\Delta t=0.0515\left(\frac{m_\nu}{E}\right)^2 D$ for a supernova
at distance $D$ (for $m_\nu$ in eV, $E$ in MeV and $D$ in kpc).  
Therefore, a spread in arrival time and correlation
of arrival time with energy can point to a non-zero neutrino mass,
assuming that the neutrinos are emitted with some energy spectrum
on a short timescale.  However, because there is
a relatively large intrinsic emission time spread ($\sim$~10 seconds),
one can in practice only get an
upper limit on $m_\nu$.  For 1987A, for reasonable assumptions about
the nature of the source time spread, 
a typical limit is $m_{\bar{\nu}_e}< 20$~eV.
For a supernova in our Galaxy and current detectors, time of flight
sensitivity to neutrino mass is
around 3~eV for $\bar{\nu}_e$~\cite{Totani};
this is not better than the best laboratory kinematic limits
on $m_{\nu_e}$~\cite{Weinheimer,Lobashev}.  

\begin{figure}[htb]
\vspace{9pt}
\includegraphics[height=2.3in]{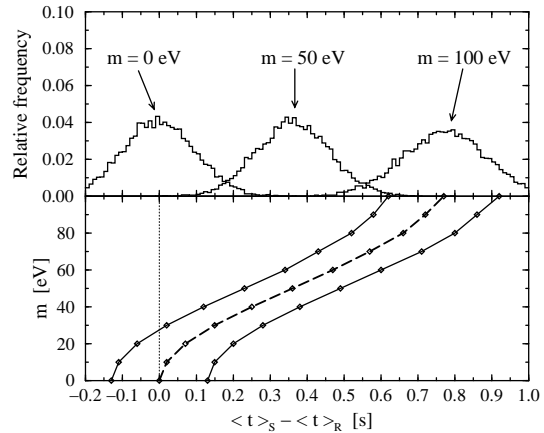}
\caption{This figure, 
from reference~\cite{Beacom}, shows the results of a study of
10000 simulated 10~kpc neutrino bursts in Super-K and SNO.  The top frame
shows distributions of relative average
time delays between NC events tagged in SNO and $\bar{\nu}_e$ events
tagged in Super-K, for three representative $\nu_{\mu,\tau}$ masses. 
The bottom frame shows the 90\% C.L. band on neutrino
mass that
one would obtain for a given measured delay.  The vertical
line at zero delay shows that one would obtain a 30~eV mass limit for
$\nu_{\mu,\tau}$ if
the NC events arrive in-time with the $\bar{\nu}_e$'s.}
\label{fig:beacom_fig}
\end{figure}

However, the next supernova may well give us mass limits for
$\nu_{\mu}$ and $\nu_{\tau}$ which are orders of magnitude better than
laboratory limits (currently 190~keV for $\nu_{\mu}$ and 15.5~MeV for
$\nu_{\tau}$~\cite{Roney}).  The idea is to use detectors with NC sensitivity to
measure relative delays between tagged NC events (which have substantial
 $\nu_{\mu,\tau}$ 
component) and $\bar{\nu}_e$, under the
assumption that the latter are light enough to
have negligible delay and can therefore provide $t=0$.
Information from more than one detector could be used. As an example,
the analysis of Beacom and Vogel~\cite{Beacom,Beacom2} (see
Figure~\ref{fig:beacom_fig}) shows that Super-K and SNO have sensitivity
to $\nu_{\mu,\tau}$ masses  in the tens of eV to keV range. 
Some more
recent work~\cite{Beacom3} shows that a supernova that continues its
collapse to a black hole could be even more interesting:
in this case, the sharp ($\sim$ sub-millisecond) cutoff in
neutrino luminosity, which should be nearly simultaneous for all
flavors, provides a clean $t=0$ for a time of flight delay measurement.  
If we are lucky enough to observe
such a supernova, the energy and time structure of the $\bar{\nu}_e$
cutoff signal could yield $\bar{\nu}_e$ mass limits rivaling laboratory ones.  In
addition, for proposed detectors such as OMNIS, the relative NC and
$\bar{\nu}_e$ time delay could give sensitivity to absolute
$\nu_{\mu,\tau}$ masses as low as 6 eV.  

\section{SNEWS}

As mentioned above, a burst of supernova neutrinos will precede the
optical signal by hours or even days; therefore, an early warning to
astronomers which could allow unprecedented early light observation
may be possible.  SNEWS (SuperNova Early Warning System) is an
international collaboration of supernova neutrino detectors, with the
goal of providing the astronomical community with a very high
confidence early alert from the coincidence of neutrino signals in several
detectors~\cite{snews2,snews3}.  In addition, the SNEWS alarm may be
able to serve as a trigger for detectors which are not able to trigger
on a supernova signal by themselves, allowing extra data to be saved.
Currently, MACRO, LVD and Super-K participate; SNO and AMANDA will be
the next active members.

Although any early warning is helpful, we would also like to be able
to tell astronomers where to look.  The question of pointing to the
supernova using the neutrino data has been examined in detail in
reference~\cite{Beacom4}.  There are two ways of pointing with
neutrinos: first, individual detectors can make use of asymmetric
reactions for which the products ``remember'' the direction of the
incoming neutrino.  Second, the timing of the neutrino signals in
several detectors can be used to triangulate; however triangulation will
be difficult with the event statistics of current detectors.  The best
bet will be to use elastic scattering interactions (equation~\ref{eq:scatt}) in
Super-K.  Estimated resolution is several degrees for a 10~kpc
supernova.

\section{SUMMARY}

In summary, several supernova neutrino detectors (Super-K, SNO, MACRO, LVD 
and AMANDA)
with sensitivity to a Galactic supernova are now running.
Others (Borexino, KamLAND, possibly OMNIS) will join them
shortly.  If a stellar core collapse occurs in our Galaxy, 
in addition to providing an early supernova alert for astronomers, these
detectors will record signals from which a wealth of physical and
astrophysical information can be mined.  

\section{ACKNOWLEDGMENTS}

The author wishes to thank all the members of the SNEWS inter-experiment
working group (in particular Alec Habig). 
John Beacom provided Figure~\ref{fig:beacom_fig}.

\end{document}